\long\global\def\C#1\F{{}}
\documentclass[11pt]{article}
\usepackage{amsmath, amsthm, amssymb, graphicx, color, verbatim, natbib} 
\swapnumbers

\newtheorem{theorem}{Theorem}
\newtheorem{lemma}[theorem]{Lemma}
\newtheorem{definition}[theorem]{Definition}

\newtheorem{remark}[theorem]{Remark}

\newtheorem{example}[theorem]{Example}

\newtheorem{corollary}[theorem]{Corollary} 
 
\newtheorem{convention}[theorem]{Convention}

\def\vtl{\vskip 1mm}
\def\tl{\vskip 2mm}
\def\wl{\vskip 4mm}

\def\too{\longrightarrow}

\def\ll{\lambda}

\def\bb{\beta}
\def\gg{\gamma}

\def\hh{\theta}

\def\Ree{\mathbb R}

\def\EQ{\Longleftrightarrow}
\def\eq{\Leftrightarrow}
\def\EOP{\phantom{a}\hfill $\square$}

\def\tooo{\overset{onto}\longrightarrow}

\def\begeq{\begin{equation}}
\def\edeq{\end{equation}}

\def\sqr#1#2{{\vcenter{\vbox{\hrule height.#2pt 
\hbox{\vrule width.#2pt height#1pt \kern#1pt 
\vrule width.#2pt}
\hrule height.#2pt}}}}
\def\square{\mathchoice\sqr56\sqr56\sqr{2.1}3\sqr{1.5}3} \def\roster{\begin{enumerate}}
\def\endroster{\end{enumerate}}

\def\fp{\noindent}

\reversemarginpar
\makeindex
\renewenvironment{proof}{{\sc Proof.}}{\EOP\wl}

\let\slantedexample\example
\def\example{\slantedexample\rm}
\let\slantedremark\remark
\def\remark{\slantedremark\rm}
\let\slanteddefinition\definition
\def\definition{\slanteddefinition\rm}

\makeatletter
\long\def\@makecaption#1#2{%
\vskip\abovecaptionskip
\sbox\@tempboxa{\small #1: \sc #2}%
\ifdim \wd\@tempboxa >\hsize
\small #1: \sc #2\par
\else
\global \@minipagefalse
\hb@xt@\hsize{\hfil\box\@tempboxa\hfil}%
\fi
\vskip\belowcaptionskip}
\makeatother

\let\realsection\section
\def\section#1{\realsection*{#1}
\addcontentsline{toc}{subsection}{#1}
\markright{#1}}

\begin{document}

\normalsize

\thispagestyle{empty}


\title{\vspace*{-1cm}
Axiomatic Derivation of the Doppler Factor\\ and Related Relativistic Laws\footnote{The authors are indebted to Janos Acz\'el, Michael Kiesling, Alexey Krioukov, David Malament and Pat Suppes for their reactions to earlier versions of the results presented here.}}
\author{Jean-Claude Falmagne\qquad\qquad\qquad\qquad\quad\,\, Jean-Paul Doignon\\[2mm] {\sl University of California, Irvine}\qquad\qquad\qquad{\sl Universit\'e Libre de Bruxelles ULB}\\[2mm] 
\hspace{1cm} jcf@uci.edu\qquad\qquad\qquad\qquad\qquad\quad\quad\, doignon@ulb.ac.be} 
\date{\today\\[4mm]
}

\def\bcint{\,]\hskip -2pt -\bb c,\bb c[\,} 
\def\gcint{\,]\hskip -2pt -\gg c,\gg c[\,} 
\def\cint{\,\,]\hskip -2pt - c, c[\,}
\def\oint{\,\,]\hskip -2pt - 1, 1[\,}
\maketitle
\vspace{-.5cm}
Lorentz-FitzGerald contraction.\\[0mm]

{\bf Running head}: Derivation of the Doppler Factor

\tl\tl

\begin{abstract} 
\fp 
The formula for the relativistic Doppler effect is investigated in the context of two compelling invariance axioms.  The axioms are expressed in terms of an abstract operation 
generalizing
the relativistic addition of velocities. We prove the following results. 
(1)~If the standard representation for the operation is not assumed \emph{a priori}, then each of the two axioms is consistent with both the relativistic Doppler effect formula and the Lorentz-Fitzgerald Contraction. 
(2)~If the standard representation for the operation is assumed, then the two axioms are equivalent to each other and to the relativistic Doppler effect formula. Thus, the axioms are inconsistent with the Lorentz-FitzGerald Contraction in this case. 
(3)~If the Lorentz-FitzGerald Contraction is assumed, then the two axioms are equivalent to each other and to a different mathematical representation for the operation which applies in the case of perpendicular motions. The relativistic Doppler effect is derived up to one positive exponent parameter (replacing the square root). 
We prove these facts under regularity and other reasonable background 
conditions. 
\end{abstract}

\thispagestyle{empty}

\section{Introduction}
A \emph{relativistic Doppler effect} arises when a source of light with wavelength $\ll$ and an observer of that source are in relative motion with respect to each other. When the source and the observer are moving towards each other at a speed $v$, the wavelength $L(\ll,v)$ perceived by the observer increases in $\ll$ and decreases in $v$, according to the special relativity formula 
\begin{align*}
\hspace{-.7cm} \text{[DE]}&&\quad\quad L(\ll,v) = \ll\,\sqrt {\frac {1-\frac vc}{1+\frac vc}}
&&\quad
\quad (\ll >0\,,\,0\leq v<c),
\end{align*}
\citep[cf.][]{ellis88,feynm63} in which
\begin{align*}
c\quad&\text{is the speed of light}\\
\ll\quad&\text{is the wavelength of the light emitted by the source}\\ L(\ll, v)\quad&\text{is the wavelength of that light measured by the observer.} 
\end{align*}

This paper analyzes the relativistic Doppler effect [DE] in the context of two compelling invariance axioms constraining the function $L$. These axioms are expressed in terms of an operation
$$
\oplus: [0,c[\,\times\,[0,c[\,\to [0,c[
$$
continuous and strictly increasing in both variables. These two axioms are: 
for any $\ll >0$, $v$, $v', w\in \,[0,c[$, 
\begin{align*}
\hskip -1cm \text {\rm [R]} &\qquad L(L(\ll ,v),w) = L(\ll , v\oplus w);\\ \noalign{\vskip .3cm}
\hskip -.85cm \text {\rm [M]} &
\qquad L(\ll ,v)\leq L(\ll ',v')
\quad\EQ\quad L(\ll ,v\oplus w)\leq L(\ll ',v'\oplus w). 
\end{align*}

If we assume that the operation $\oplus$ satisfies the standard representation for the relativistic addition of velocities, that is 
$$
\hskip -5cm \text{[AV]}\qquad\quad\qquad\quad \qquad\quad v\oplus w= \frac{v+ w}{1 + \frac{vw}{c^2}}, $$
these conditions are easy to interpret and intuitively cogent. We derive three sets of equivalences. In our first main result in Theorem~\ref{gen theo}, we do not assume \emph{a priori} that the operation $\oplus$ has the form [AV]. Under natural background conditions on the function $L$, we prove that [R] and [M] are equivalent, and that each of them is equivalent to the following generalizations of [DE] and [AV] in which $\xi$ is a positive constant and $u$ is a continuous strictly increasing function mapping the interval $[0,c[$ onto itself: 
\begin{roster}
\item[{[D$^\dagger$]}]\qquad\quad$L(\ll , v) = \ll \left (\frac{c-u\left( v\right)}{c+u\left(v\right)}\right)^{\xi}$, 
\item[{[A$^\dagger$]}]\qquad\quad$v\oplus w = u^{-1}\left(\frac {u\left(v\right) +u\left( w\right)}{1 + \frac{u\left( v\right)u\left(w\right)}{c^2}}\right)$. 
\end{roster}
Thus, [DE] and [AV] obtain when $u$ is the identity function and $\xi = \frac 12$. Specifically, our Theorem \ref{gen theo} establishes the equivalences \wl
\centerline{[R]\quad $\EQ$\quad \big([D$^\dagger$] $\&$ [A$^\dagger$]\big)\quad $\EQ$\quad [M].} \wl

Theorem \ref {gen theo} has a two corollaries. In Corollary \ref{coro}, we assume that the operation $\oplus$ satisfies the standard formula [AV] for the relativistic addition of velocities and we derive the stronger result: 
\wl
\centerline{[R]\quad $\EQ$\quad [D$^\star$]\quad $\EQ$\quad [M],} \wl

\fp
with, for some positive constant $\xi$,
\begin{roster}
\item[\,\,{[D$^\star$]}] \quad\quad $L(\ll,v) = \ll \left(\frac {1-\frac vc}{1+\frac vc}\right)^\xi$. \end{roster}

Axioms [R] and [M] are thus inconsistent with the Lorentz-FitzGerald Contraction if $\oplus$ satisfies [AV]. In our second corollary, we do not assume that $\oplus$ satisfies [AV]. Instead, we assume that [LF] holds and we show that each of [R] and [M] is equivalent to the representation 
$$
\text{\rm[A$^\star$]}\quad\,\,\, v\oplus w= c\, \sqrt{{{\left(\frac vc\right)^2+\left(\frac wc\right)^2- \left(\frac wc\right)^2\left(\frac vc\right)^2}}}\qquad\quad(v,w\in [0,c[\,), 
$$
which applies in the case of perpendicular motions \citep[see e.g.][Eq.~(8)]{ungar91}.
We prove thus that
\wl
\centerline{[LF] \quad$\Longrightarrow$\quad([R]\,\, $\EQ$\,\, [A$^\star$] \,\, $\EQ$\quad [M]).} 
\wl
All the results are established under natural regularity and other background conditions on the function $L$ (see Convention \ref{convention}). 

\section{Three Preparatory Lemmas}

We write $\Ree_+$ for the set of (strictly) positive real numbers. The following result is well-known\footnote{The quoted result from \citet{aczel:66} requires that $x,y\in\,\, ]-1,1[$ but the functional equation (\ref{xy=rel x+y}) can be extended to $\widetilde m$ on $]-1,+1[$ by defining $\widetilde m(-x) = \frac 1{m(x)}$.} \citep[see][]{aczel:66}.

\begin{lemma}\label{relat addition} 
The set of solutions $m:[0,1[\, \to \Ree$ of the functional equation 
\begin{equation}\label{xy=rel x+y}
m\left (\frac {x+y}{1+xy}\right )
=
m(x)\,m(y) 
\quad\qquad(x,y\in [0,1[\,\,), 
\end{equation}
with $m$ strictly decreasing, is defined by the equation 
\begin{equation}\label{sol m}
m(x) = \left (\frac {1-x}{1+x}\right )^\xi\quad\qquad(x\in [0,1[\,\,),
\end{equation}
where $\xi\in\Ree_+$.
\end{lemma}

We nevertheless include a proof.
\tl
{\sc Proof.} The operation $ (x,y)\mapsto \frac {x+y}{1+xy}$ on $\,]\hskip -2pt-1,1[$ is  isomorphic to
$(\Ree_+,\cdot)$.
Specifically, with
\begin{equation}\label{def tanh}
\tanh (x) = \frac { e^x - e^{-x}}{e^x + e^{-x}}\quad\qquad(x\in\Ree_+), 
\end{equation}
and
\begin{equation}\label{def s t}
x = (\tanh\circ \ln)( s), \quad \text{and}\quad y = (\tanh\circ \ln)( t), 
\end{equation}
we have, for $s,t\in\Ree_+$,
\begin{equation}\label{isom}
(\tanh\circ \ln)(st) = \frac { (\tanh\circ \ln)( s) + (\tanh\circ \ln)( t)} {1 + (\tanh\circ \ln)( s) \, (\tanh\circ \ln)( t)}\,. \end{equation}
Defining
\begin{equation}\label{def r}
F = m\circ\tanh\circ \ln,
\end{equation}
and using (\ref{def s t}) and (\ref{isom}), we can rewrite (\ref{xy=rel x+y}) in the form of the Cauchy-type equation
$$
F(st) = F(s)\,F(t),\quad\qquad (s,t\in \Ree_+).
$$
Because $m$ is strictly decreasing, so is the function $F = m\circ\tanh\circ \ln$ and we obtain by a standard functional equation result \citep[cf.][]{aczel:66} 
\begin{equation}\label{r=}
F(s) = s^{-\ll}\quad\qquad(s\in\Ree_+)
\end{equation}
for some $\ll\in\Ree_+$.
Equation (\ref{sol m}) is easily derived from (\ref{def r}) and (\ref{r=}). \EOP

\def\minop{\,]\hspace{-3pt}-\hspace{-3pt}} \begin{lemma}\label{ff = fo}
Let $\oplus$ be a real operation on an open interval interval $[\,0,c\,[\,$, with $c\in\Ree_+$ and having $0$ as its identity element. Suppose that there is some continuous strictly decreasing function $f$ mapping $[\hspace{1 pt}0,c\,[\,$ onto $\,]\hspace{1 pt}0,1]$ such that 
\begin{equation}\label{basic ff = f}
f(v\oplus w) = f(v)f(w)\quad\qquad(v,w\in [0,c\,[\,).
\end{equation}
Then there exists a continuous strictly increasing function $u$ mapping
 $[\hspace{1 pt}0,c[$ onto $[0,c[$ and a constant $\xi >0$ such that \begin{equation}\label{basic f = c-u/c+u} f(v) = \left
(\frac{c-u\left( v\right)}{c+u\left(v\right)}\right)^{\xi}\qquad\quad(v\in [0,c\,[\,). \end{equation}
\end{lemma}

\begin{proof} Equation (\ref{basic ff = f}) implies that the function $f$ is an isomorphism of $\oplus$ onto the restriction of the multiplicative group $(\Ree_+,\cdot)$ of the positive reals to the interval $]0,1]$. But,
as in the proof of Lemma~\ref{xy=rel x+y},
$(\Ree_+,\cdot)$ is isomorphic to the group $G=(\minop 1,1[, \odot)$, with the operation $\odot$ defined by $$
x\odot y= \frac {x+ y} {1+xy}.
$$
Thus, $\oplus$ is isomorphic to the restriction of $G$ to the interval $[0,1[$. Accordingly, there exists a strictly increasing continuous bijection $g:[0,1[\,\to [0,c[\,$ such that 
$$
g\left(\frac{x+ y}{1+xy}\right) = g(x) \oplus g(y) \qquad\qquad(x,y\in [0,1[\,). 
$$
Applying the function $f$ on both sides, we get 
\begin{align*}
(f\circ g)\left(\frac{x+ y}{1+xy}\right) &= f\left (g(x) \oplus g(y)\right)&&\\ &= f(g(x)) f(g(y))&&(\text{by(\ref{basic ff = f})}), 
\end{align*}
or with $m(x) =(f \circ g)(x)$ for $x \in[0,1[$, $$
m\left(\frac{x+y}{1+xy}\right) = m(x)m(y)\qquad\qquad(x,y\in[0,1[\,). $$
Using Lemma \ref{relat addition}, we obtain \begin{equation}\label{fg = m =}
(f \circ g)(x)= m(x) = \left (\frac{1-x}{1+x}\right)^\xi=\left (\frac{c-cx}{c+cx}\right)^\xi\qquad\qquad(x\in[0,1[\,). \end{equation}
For any $x\in [0,1\,[$, we have $g(x) = v$ for some $v\in [0,c\,[$. We now define $u(v) = cg^{-1}(v)$. Note that, as  $g^{-1} :[0,1[\,\to[0,1[$ is a continuous bijection, the function $u: [0,c[\,\to [0,c[$ is a continuous strictly increasing bijection. We can thus rewrite (\ref{fg = m =}) as
$$
f(v) = \left
(\frac{c-u\left( v\right)}{c+u\left(v\right)}\right)^{\xi}\qquad\quad(v\in [0,c\,[\,), $$
establishing the lemma.
\end{proof}

\begin{definition}\label{left o-i} A function $H:\Ree_+\times [0,c[\,\to \Ree_+$ is \emph{left order-invariant with respect to similarity transformations} if for any $x,z\in \Ree_+$, $a >0$, and $y,w\in [0,c[$, we have \begin{align*}
&\text{\rm [LOI]}
\,\, &&H(x,y)\leq H(z,w)\quad\EQ\quad H(ax,y)\leq H(az,w).\\[2mm] \noalign{In the sequel, we simply say in such a case that $H$ satisfies [LOI]. The function $H$ is said to satisfy the \emph{double cancellation} condition if, for all $x,z,t\in\Ree_+$ and $y,w, s\in [0,c[$ we have 
\vspace*{4mm}} 
&\text{\rm [DC]} && H(x,y) \leq H (z,w)\,\, \&\,\, H (z, s) \leq H( t, y)\, \Rightarrow\,H (x,s)\leq H ( t,w).
\end{align*}
\end{definition}

The double cancellation condition appears in \citet[][and related references]{krant71}; it is satisfied for instance by the usual addition of real numbers.

\begin{lemma}\label{ll > ll' => all > all' }Suppose that a function $L: \Ree_+\times [0,c[\,\,\overset{onto}\too \Ree_+$ is strictly increasing in the first variable, strictly decreasing in the second variable, and continuous in both. \begin{roster}
\item[{\rm (i)}]
 $L$ satisfies Condition {\rm [LOI]} if and only if there exist two functions\,\,
$F:\Ree_+\to\Ree_+$ and\linebreak $f:[0,c[\,\,\overset{onto}\too\,\, ]0,1]$, respectively strictly increasing and strictly decreasing, such that $f(0) = 1$ and 
$$
L(\ll, v) = F\left(\ll f\left( v\right)\right). 
$$
\item[{\rm (ii)}] If we also suppose that $L(\ll, 0) = \ll$ for all $\ll\in [0,\infty[$, then 
\begin{equation}\label{(ii) in Lem 4}
L(\ll, v) =\ll f\left( v\right)
\end{equation}
 with $f$ continuous.
\end{roster}
\end{lemma}

These results are byproducts of Theorems 21 and 38 in \citet{falma04}. We include a proof for completeness. 
\vtl
\begin{proof} First notice that [LOI] is satisfied when there exist functions $F$ and $f$ as in the statement.  Next,  suppose that a function $L$ satisfies the stated background conditions and [LOI]. For any constant $a\in\Ree_+$, define the function $\phi_a:\Ree_+\to~\Ree_+$ by the equation 
\begin{equation}\label{def phi_a}
(\phi_a\circ L)(\ll,v) = L(a\ll, v)\qquad\qquad(\ll \in\Ree_+, v\in [0,c[\,). 
\end{equation}
The [LOI] condition implies that the function $\phi_a$ is well defined and increasing for any $a\in\Ree_+$. It is easily verified that any two functions $\phi_a$ and $\phi_b$ commute in the sense that $\phi_a\circ\phi_b\circ L = \phi_b\circ\phi_a\circ L$. We begin by showing that, under the hypotheses of the lemma, 
the function $L$ must satisfy the double cancellation condition [DC] introduced in Definition \ref{left o-i}. Suppose that, for some $\ll$, $\zeta$ and $\hh$ in $\Ree_+$ and $v$, $w$ and $s$ in $[0,c[$, we have \begin{equation}\label{doub canc hyp}
L (\ll, v) \leq L (\zeta,w)\,\, \&\,\, L (\zeta, s) \leq L( \hh, v). \end{equation}
Define $a =\frac \hh\zeta$ and $b = \frac \ll \hh$. We have successively
\begin{align*}
L( \hh, w) = L (a\zeta,w)&=( \phi_a\circ L)(\zeta,w)&&(\text{by (\ref{def phi_a})})\\ &\geq ( \phi_a\circ L)(\ll,v)&&(\text{by (\ref{doub canc hyp})})\\ &= ( \phi_a\circ L)(b\hh,v)&&\\
&= ( \phi_a\circ\phi_b\circ L)(\hh,v)&&(\text{by (\ref{def phi_a})})\\ &\geq ( \phi_a\circ\phi_b\circ L)(\zeta,s)&&(\text{by (\ref{doub canc hyp})})\\ &= ( \phi_b\circ\phi_a\circ L)(\zeta,s)&&(\text{by commutativity})\\ &= ( \phi_b\circ L)(a\zeta,s)&&(\text{by (\ref{def phi_a})})\\ &= ( \phi_b\circ L)(\hh,s)&&\\
&= L(b\hh,s)&&(\text{by (\ref{def phi_a})}).\\ &= L(\ll,s)&&
\end{align*}
Thus, (\ref{doub canc hyp}) implies $L( \hh, w) \geq L(\ll,s)$. So, the function $L$ satisfies [DC]. By a standard result of measurement theory \citep[see][Theorem 2, p.~257]{krant71}, 
recast in the context of real variables and the continuity and strict monotonicity assumptions on the function $L$, this implies that there exists three functions 
$k: \Ree_+ \to \Ree_+$,
$h:[0,c[\, \to \Ree_+\,$ and $G: [0,\infty[\,\to [0,\infty[$, with $k$ and $G$ strictly increasing and $h$ strictly decreasing, satisfying \begin{equation}\label {x representation} L(\ll,v) = G(k(\ll)h(v)).
\end{equation}
From (\ref{x representation}), we infer that 
\begin{gather}\label {phi_a G}
(\phi_a\circ G) (k(\ll)h(v))= G(k(a\ll)h(v)) \quad\quad(a, \ll\in \Ree_+, v\in[0,c[\,). \end{gather}
We now rewrite the function $\phi_a$ in terms of $k$, $G$ and $a$. Redefining the function $k$ if need be, we can assume that \begin{equation}\label{h(0) = 1}
h(0) = 1.
\end{equation}
Thus, with $v = 0$, (\ref {phi_a G}) becomes $$
(\phi_a\circ G) (k(\ll))= G(k(a\ll)).
$$
Setting $t= (G\circ k)(\ll)$, we get, with $\ll = (k^{-1}\circ G^{-1}) (t)$ \begin{equation}
\phi_a(t) = (G\circ k\circ ak^{-1}\circ G^{-1}) (t), \end{equation}
which defines the function $\phi_a$ in terms of the functions $G$, $k$ and the constant $a$. Substituting $\phi_a$ in (\ref{phi_a G}) by its expression given by (\ref{def phi_a}), we obtain $$
(G\circ k\circ ak^{-1}\circ G^{-1}\circ G)(k(\ll)h(v)) = G(k(a\ll)h(v)). $$
Applying the function $G^{-1}$ on both sides and simplifying, we get \begin{equation}\label{simplify phi}
(k\circ ak^{-1})(k(\ll)h(v)) = k(a\ll)h(v). 
\end{equation}
Defining now $x = k(\ll)$ and $y = h(v)$ and $g_a = k\circ ak^{-1}$, Eq.~(\ref{simplify phi}) can be rewritten as $
g_a(xy) = g_a(x)y,
$
So, the function $g_a$ is necessarily of the form $$
g_a(x) = C(a) x.
$$
for some function $C: \Ree_+ \to \Ree_+$. We have thus $
(k\circ ak^{-1}) (x) = C(a) x,
$
that is
\begin{equation}\label{pexider}
k(a \ll) = C(a) k(\ll),
\end{equation}
a Pexider equation defined for all $a$ and $\ll$ in $\Ree_+$. Because the function $k$ is strictly increasing, the set of solutions of (\ref{pexider}) is given by the equations: \begin{align}\label{def k}
k(\ll) &= A\ll^B\\
\label{def C}
C(a) &= a^B,
\end{align}
with positive constants $A$ and $B$ \citep[cf.][Theorem 4, p.~144]{aczel:66}. We obtain thus \begin{equation}\label{G(A ll B h(v))}
L(\ll,v) = G(A\ll^B h(v))\qquad\qquad(\ll\in\Ree_+\,, v\in [0,c[\,). \end{equation}
Defining the functions
\begin{align}\label{def f}
f(v) &= h(v)^{\frac 1B}&&(x\in[0,1[\,)\\ \label{def F}
F(t) &= G(At^B)&&(t\in[0,\infty[\,),
\end{align}
we obtain
\begin{equation}\label{result for (i)}
L(\ll,v) = G(A\ll^B h(v)) = G\left( A \left (\ll h(v)^{\frac 1B}\right)^B\right) = F\left(\ll f\left(v\right)\right). \end{equation}
Finally, by (\ref{h(0) = 1}), we have $f(0) = h(0)^{\frac 1B} = 1$. (We recall that the function $k$ of (\ref{x representation}) may have been redefined to ensure that $h(0) = 1$. This redefinition of $k$ is thus affecting the constant $A$ of (\ref{def k}), which is now absorbed by the function $F$.) Thus $f$ maps $[0,1[$ onto $]0,1[$. This proves (i). \tl

Condition (ii) results immediately from (\ref{result for (i)}) since $$
\ll = L(\ll, 0) = F(\ll f(0)) = F(\ll),
$$
and so $F$ is the identity function. Since  $F$ is continuous in its second variable,  the function $f$ in (\ref{(ii) in Lem 4}) is continuous.
\end{proof}
\section{Main Results}

\begin{convention}\label{convention} {\rm In the sequel, when we write that some function $L: \Ree_+\times [0,c[\,\,\overset{onto}\too \Ree_+$ satisfies the \emph{basic hypotheses}, we mean that $L$ is strictly increasing in the first variable, strictly decreasing in the second variable, continuous in both, 
and satisfies [LOI] plus the additional conditions
of Lemma \ref{ll > ll' => all > all' }.
Accordingly, there exists by that lemma a continuous strictly decreasing function $f:[0,c[\,\,\overset{onto}\too\,\, ]0,1]$ such that \begin{equation}\label{L = l x f}
L(\ll, v) =\ll f\left( v\right).
\end{equation}
}
\end{convention}
For convenience of reference, we reproduce the formulas of our two main axioms. For $\ll,\ll'\in\Ree_+$ and $ v,v',w\in [0,c[\,$:
 \begin{align*}
\hskip -1cm \text {\rm [R]} &\qquad L(L(\ll ,v),w) = L(\ll , v\oplus w);\\ \noalign{\vskip .3cm}
\hskip -.85cm \text {\rm [M]} &
\qquad L(\ll ,v)\leq L(\ll ',v')
\quad\EQ\quad L(\ll ,v\oplus w)\leq L(\ll ',v'\oplus w). \end{align*}

The theorem below concerns the representations of the function $L$ and the operation~$\oplus$ in terms of a continuous, strictly increasing function $u:[0,c[\,\,\tooo [0,c[\,$ and a positive constant $\xi$. We also recall the two representations: for $v,w\in [0,c[\,$: \begin{roster}
\item[{[D$^\dagger$]}]\qquad\quad $L(\ll , v) = \ll \left (\frac{c-u\left( v\right)}{c+u\left(v\right)}\right)^{\xi}$; \item[{[A$^\dagger$]}]\qquad\quad\, $v\oplus w = u^{-1}\left(\frac {u\left(v\right) +u\left( w\right)}{1 + \frac{u\left( v\right)u\left(w\right)}{c^2}}\right).$ \end{roster}

\begin{theorem}\label{gen theo} Suppose that a function $L: \Ree_+\times [0,c[\,\,\overset{onto}\too \Ree_+$ satisfies the basic hypotheses, and let $$
\oplus: [0,c[\,\times[0,c[\,\to [0,c[\qquad\quad(c >0) $$
be an operation having $0$ as its identity element. We assume that $\oplus$ is continuous and strictly increasing in both variables. Under those hypotheses, the following equivalences hold: \wl
\centerline{\rm [R]\quad $\EQ$\quad \big([D$^\dagger$] $\&$ [A$^\dagger$]\big)\quad $\EQ$\quad [M].} \wl
Moreover, the same function $u$ is involved in both {\rm [D$^\dagger$]} and {\rm [A$^\dagger$]}. \end{theorem}
\tl

\begin{proof} We successively prove [R] $\Rightarrow$ ([D$^\dagger$] $\&$ [A$^\dagger$]) $\Rightarrow$ [M] $\Rightarrow$ [R]. \tl

[R] $\Rightarrow$ [D$^\dagger$].
By Lemma \ref{ll > ll' => all > all' } and Axiom [R], we get 
\begin{equation*}
\ll f(v\oplus w) = L(\ll,v\oplus w) = L(L(\ll, v), w) = L(\ll, v)f(w)
=\ll f(v)f(w),
\end{equation*}
and so, with $f:[0,c[\,\,\overset{onto}\too\,\,
]0,1]$
strictly decreasing and continuous, 
\begin{equation}\label{ff = f(oplus)}
f(v\oplus w) = f(v)f(w)\qquad\qquad(v,w\in[0,c[). 
\end{equation}

Applying Lemma \ref{ff = fo} gives
\begin{equation}\label{gen f with u}
f(v)= \left
(\frac{c-u\left( v\right)}{c+u\left(v\right)}\right)^{\xi} \end{equation}
for some continuous, strictly increasing function $u: [0,c[\,\,\overset{onto}\too\,\,[0,c[$, and some constant $\xi >0$. So [D$^\dagger$] holds. 

\wl

[R] $\Rightarrow$ [A$^\dagger$]. From (\ref{gen f with u}), we get
for $z\in\,]0,1]$ 
\begin{equation}\label{f-1}
f^{-1}(z)  = u^{-1}\left(c\times \frac{ 1- z^{1/\xi}} {1 + z^{1/\xi}}\right). \end{equation}
Using now (\ref{ff = f(oplus)}), (\ref{gen f with u}) and (\ref{f-1}), we obtain \begin{align}\nonumber
v\oplus w&= f^{-1}\left(f(v)f(w)\right)\\ \label{oplus = f-1 ff}
& = f^{-1}\left(
\left(\frac{c-u\left( v\right)}{c+u\left ( v\right)}\right)^\xi 
\left(\frac{c-u\left( w\right)}{c+u\left ( w\right)}\right)^\xi \right)\\
\label{oplus = f-1 ff 2}
&= u^{-1}\left( c\times \frac{1-\left(\frac{ c -u\left( v\right)}{ c +u\left( v\right)}\right)\left(\frac{ c -u\left( w\right)}{ c +u\left( w\right)} \right)}{1+\left(\frac{ c -u\left( v\right)}{ c +u\left( v\right)}\right)\left(\frac{ c -u\left( w\right)}{ c +u\left( w\right)} \right)} \right)\\ \label{oplus = f-1 ff 3}
&=u^{-1}\left(\frac {u\left(v\right) +u\left( w\right)}{1 + \frac{u\left( v\right)u\left(w\right)}{ c ^2}}\right), \end{align}
after simplifications. We have thus proved that [R] $\Rightarrow$ ([D$^\dagger$] $\&$ [A$^\dagger$]). \tl

([D$^\dagger$] $\&$ [A$^\dagger$]) $\Rightarrow$ [M]. In the derivation below, we use the abbreviation 
$$
f(v) = \left(\frac{ c -u\left( v\right)}{c+u\left(v\right)}\right)^{\xi}. 
$$
By [A$^\dagger$] and 
calculations similar to
(\ref{oplus = f-1 ff})-(\ref{oplus = f-1 ff 3}) we
derive that 
\begin{equation}\label{use [A dagger]}
f(v\oplus w) = f(v)f(w).
\end{equation}
For any $\ll,\ll'\in \Ree_+$ and $v,v',w\in [0,c[$, we now have 
\begin{align*}
L(\ll, v) \leq L(\ll',v') &\EQ \ll f(v)\leq \ll' f(v') &&(\text{by [D$^\dagger$]})\\ 
&\EQ \ll f(v)f(w)\leq \ll' f(v')f(w) &&(\text{since $f(w) > 0$})\\ &\EQ \ll f(v\oplus w)\leq \ll' f(v'\oplus w) &&(\text{by (\ref{use [A dagger]})})\\ &\EQ L(\ll,v\oplus w)\leq L(\ll',v\oplus w)&&(\text{by [D$^\dagger$]}). \end{align*}
and thus [M] holds.

\tl

[M] $\Rightarrow$ [R]. An immediate consequence of [M] is that there exists, for any $w\in [0,c[$, a function $K_w: \Ree_+\to \Ree_+$ such that
\begin{equation}\label{def Kw}
K_w(L(\ll,v)) = L(\ll,v\oplus w)\qquad\quad(\ll\in\Ree_+, v\in [0,c[\,). \end{equation}
(Axiom [M] ensures that $K_w$ is well defined by the above equation.). Applying 
the basic hypotheses and Lemma~\ref{ll > ll' => all > all' }, Eq.~(\ref{def Kw}) becomes \begin{equation}\label{Kw}
K_w(\ll f(v)) = \ll f(v\oplus w).
\end{equation}
Setting $v=0$, we obtain $0\oplus w =w$ in the r.h.s.~of (\ref{Kw}), and since $f(0) = 1$ by Lemma \ref{ll > ll' => all > all' }, we get $$
K_w(\ll) = \ll f(w).
$$
This allows us to rewrite (\ref{Kw}) as
$
\ll f(v)f(w) = \ll f(v\oplus w),
$
or equivalently
$$
L(L(\ll,v),w) = L(\ll, v\oplus w).
$$
Thus, [R] holds. This completes the proof of the theorem. \end{proof}
In the next corollary, we assume that the operation $\oplus$ satisfies the standard formula for the relativistic addition of velocities: $$
\text{[AV]}\qquad\quad\qquad\quad \qquad\quad v\oplus w= \frac{v+ w}{1 + \frac{vw}{c^2}}\qquad\quad(v,w\in [0,c[\,). $$
As a result, we obtain a form for the function $L$ which only differs from the Doppler operator by a unspecified positive exponent $\xi$. \begin{corollary}\label{coro} Suppose that a function $L: \Ree_+\times [0,c[\,\,\overset{onto}\too \Ree_+$ satisfies the basic hypotheses, and let now $\oplus$ be defined by the standard formula {\rm [AV]} for the relativistic addition of velocities. Then the following three equivalences hold: 

\vtl

\centerline{\rm [M]\quad $\EQ$\quad [D$^\star$]\quad $\EQ$\quad [R],} \tl
\hspace{-.7cm} with, for some exponent $\xi\in\Ree_+$, 
$$
\text{\rm[D$^\star$]}\quad\quad\quad\quad \qquad\quad L(\ll,v) = \left(\frac {1-\frac vc}{1+\frac vc}\right)^\xi\qquad\quad(v,w\in [0,c[\,). $$
\end{corollary}

\begin{proof} As [D$^\star$] is a special case of [D$^\dagger$] and [AV] implies [A$^\dagger$] , we know by Theorem~\ref{gen theo} that [D$^\star$] $\Rightarrow$ [M] $\eq$ [R]. It suffices thus to establish that [M] $\Rightarrow$ [D$^\star$]. We only sketch the argument (which is familiar from our proof of Theorem \ref{gen theo}). Axiom [M] implies that \begin{equation}\label{M -> Ds1}
K_w(L(\ll,v)) = L(\ll,v\oplus w)\qquad\quad(\ll\in\Ree_+, v\in [0,c[\,) \end{equation}
for some function $K_w$. By Lemma \ref{ll > ll' => all > all' }, we get $
K_w(\ll f(v)) = \ll f(v\oplus w).
$
Setting $v=0$ yields $K_{w}(\ll ) = \ll \,f(w)$. We can thus rewrite (\ref{M -> Ds1}), after cancelling the $\ll $'s, as \begin{equation}\label{M -> Ds2}
f(v)f(w)=
f\left (\frac{v + w}{1 + \frac{vw}{c^2}}\right )\,.
\end{equation}
Defining $m:[0,1[ \to [0,1[\, : x \mapsto m(x) = f(c x)$, we obtain
\begin{equation}\label{xy=rel x+y 2}
m(x)\,m(y) = m\left (\frac {x+y}{1+ xy}\right )\quad\qquad(x,y\in [0,1[\,), \end{equation}
with $m$ strictly decreasing and continuous. Applying Lemma \ref{relat addition} yields \begin{equation}\label{xy=rel x+y 3}
f(cx)= m(x) = \left (\frac {1-x}{1+x}\right )^\xi \end{equation}
for some constant $\xi >0$. With $v=cx$, Lemma \ref{ll > ll' => all > all' } and (\ref{xy=rel x+y 3}) give \begin{equation}\label{L=}
L(\ll ,v) = \ll f(v) =\ll  \left
(\frac{1-\frac v{c}}{1+\frac v{c}}\right)^{\xi}\,\quad\qquad (v\in [0,c[\,). \end{equation}
Thus, [D$^\star$] holds.
\end{proof}

\begin{remark}{\rm Corollary \ref{coro} tells us that if the operator $\oplus$ is defined by [AV], then Axioms [R] and [M] are equivalent to~[D$^*$]. However, whatever the value of its exponent $\xi> 0$, Formula [D$^*$] is inconsistent with the Lorentz-FitzGerald Contraction \begin{roster}
\item[]
\begin{roster}
\item[{[LF]}]\quad\qquad $L(\ll,v) = \ll \sqrt{1-\left(\frac vc\right)^2}$. \end{roster}
\end{roster}
The point is that [LF] and [D$^*$] represent essentially different functions for any fixed positive value of the exponent $\xi$ in [D$^*$]. Thus, if the standard formula for the addition of velocities is assumed, then the Lorentz-FitzGerald Contraction is inconsistent with our two axioms [R] and [M], a fact which may be worth pondering. }
\end{remark}
\tl
On the other hand, if we do not assume {\sl a priori} that the operation $\oplus$ is defined by the standard formula [AV], then the Generalized Doppler Formula [D$^\dagger$] {\bf is} consistent with the Lorentz-FitzGerald operator in the sense that the equation $$
\sqrt{\frac {c-u(v)}{c+u(v)}} = \sqrt{1-\left(\frac vc\right)^2} $$
can be solved for the function $u$.
The solution is
\begin{equation}\label{Dop-LF}
u(v)=\frac{c\left(\frac vc\right)^2}{2 -\left(\frac vc\right)^2}. \end{equation}
This raises the question: what is the form of an operation $\oplus$ that is consistent with [R], [M] and the Lorentz-Fitzgerald Contraction? Our next corollary answers the question. 

\begin{corollary} \label{coro 2}
Suppose that a function $L: \Ree_+\times [0,c[\,\,\overset{onto}\too \Ree_+$ satisfies the basic hypotheses. Let $$
\oplus: [0,c[\,\times[0,c[\,\to [0,c[\qquad\quad(c >0) $$
be an operation that is continuous and strictly increasing in both variables and has $0$ as its identity element. If the function $L$ satisfies the Lorentz-FitzGerald Contraction {\rm [LF]}, then the following three equivalences hold: \vtl
\centerline{\rm [M]\quad $\EQ$\quad [A$^\star$]\quad $\EQ$\quad [R],} \tl
\hspace{-.7cm} with
$$
\text{\rm[A$^\star$]}\quad\qquad\,\,\, v\oplus w= c\, \sqrt{{{\left(\frac vc\right)^2+\left(\frac wc\right)^2- \left(\frac wc\right)^2\left(\frac vc\right)^2}}}\qquad\quad(v,w\in [0,c[\,). 
$$
\end{corollary}

Note that, for $ v,w\in [0,c[\,$, [A$^\star$] can be rewritten as
 \begin{align*}
v\oplus w&= c\, \sqrt{{{\left(\frac vc\right)^2\left (1 -\left(\frac wc\right)^2\right) + \left(\frac wc\right)^2}}}\\ &= c\, \sqrt{{{\left(\frac wc\right)^2\left (1 -\left(\frac vc\right)^2\right) + \left(\frac vc\right)^2}}}, \end{align*}
so that $\oplus$ is strictly increasing in both variables and also satisfies \begin{gather*}
0\leq v\oplus w < c,\\
\lim_{v\to c}v\oplus w = \lim_{w\to c}v\oplus w=c,\\ \noalign{and}
0\oplus v= v\oplus 0 = v.
\end{gather*}

\begin{proof}
We first show that [LF] and [A$^\star$] are special cases of [D$^\dagger$] and [A$^\dagger$], respectively. Beginning with [LF], we have by [D$^\dagger$] , with $\xi = \frac 12$, 
\begin{equation*}
L(\ll,v)=\ll\sqrt{\frac{c-u(v)}{c+u(v)}}\quad\quad(\ll\in \Ree_+,\, v\in [0,c[\,). \end{equation*}
Replacing $u(v)$ by its expression in (\ref{Dop-LF}) gives [LF]. 

We prove that [A$^\star$] is a special case of [A$^\dagger$] the same way, that is, by replacing the function $u$ in [D$^\dagger$] by its expression in (\ref{Dop-LF})---thus $$
u^{-1}(t) = c\,\sqrt{\frac{2t}{c+t}}.
$$
This involves lengthier manipulations, which we omit. We obtain: \begin{equation*}
v\oplus w = u^{-1}\left(\frac {u\left(v\right) +u\left( w\right)}{1 + \frac{u\left( v\right)u\left(w\right)}{c^2}}\right) = c\, \sqrt{{{\left(\frac vc\right)^2+\left(\frac wc\right)^2- \left(\frac wc\right)^2\left(\frac vc\right)^2}}}. \end{equation*}
By Theorem \ref{gen theo}, we have [R] $\eq$ ([D$^\dagger$] $\&$ [A$^\dagger$]) $\eq$ [M]. So far, we have thus established that 
$$
\text{[LF]}\quad\Longrightarrow\quad\big(\text{[A$^\star$]}\,\,\Rightarrow\,\,\text{[R}]\,\,\land\,\,\text{[M]}\big).
$$
In fact, it is easily shown that
\begin{equation}\label{LF => ([Adag] => [R] <=> [M]} 
\text{[LF]}\quad\Longrightarrow\quad\big(\text{[A$^\star$]}\,\,\Rightarrow\,\,(\text{[R}]\,\,\eq\,\,\text{[M]})\big). 
\end{equation}
Suppose now that both [LF] and [R] hold. We have thus \begin{align*}
L(L(\ll,v),w) &= \ll \sqrt{1-\left(\frac vc\right)^2} \sqrt{1-\left(\frac wc\right)^2}\\ &= \ll \sqrt{1-\left(\frac {v\oplus w}c\right)^2}, \end{align*}
which implies
$$
\left(1-\left(\frac vc\right)^2\right)\left(1-\left(\frac wc\right)^2\right) = 1 - \left(\frac {v\oplus w}c\right)^2. $$
Solving for $v\oplus w$ yields  [A$^\star$]. We have thus also \begin{equation}\label{LF => ( [R] <=> [M] => [Adag]} \text{[LF]}\quad\Longrightarrow\quad(\text{[R]}\,\,\Rightarrow\,\,\text{[A$^\star$]}). \end{equation}
Combining (\ref{LF => ([Adag] => [R] <=> [M]}) and (\ref{LF => ( [R] <=> [M] => [Adag]}) gives the equivalences of the theorem. \end{proof}

\subsubsection*{\Large References}

\begin{itemize}
\bibitem[Acz\'el(1966)]{aczel:66}
\hspace{-1cm} Acz\'el, J. (1966).
\newblock \emph{Lectures on Functional Equations and their Applications}. \newblock Academic Press, New York.

\bibitem[Ellis and Williams(1966)]{ellis88} \hspace{-1cm} Ellis, G.F.R. and Williams, R.M. (1988). \newblock \emph{Flat and Curved Space-Times}. \newblock Clarendon Press, Oxford.

\bibitem[Feynman et~al.(1963)Feynman, Leighton, and Sands]{feynm63} \hspace{-1cm} Feynman, R.P., Leighton, R.B. and Sands, M. (1963). \newblock \emph{The {F}eynman Lectures on Physics}. \newblock Addison-Wesley, Reading, MA.

\bibitem[Falmagne(2004)]{falma04}
\hspace{-1cm} Falmagne, J.-Cl.
\newblock Meaningfulness and order invariance: two fundamental principles for scientific 
laws.
\newblock \emph{Foundations of Physics}, 9:\penalty0 1341--1384, 2004. 

\bibitem[Krantz, Luce, Suppes and Tversky(1971)]{krant71} \hspace{-1cm} Krantz, D.H., Luce, R.D., Suppes, P. and Tversky, A. (1971) \newblock \emph{Foundations of Measurement, Volume 1: Additive and Polynomial 
Representations}.
\newblock Academic Press, New York and San Diego. 

\bibitem[Ungar(1991)]{ungar91}
\hspace{-1cm} Ungar, A.A.
\newblock Thomas precession and its associated grouplike structure.
\newblock \emph{American Journal of Physics}, 59(9):\penalty0 824--834, 1991. 
\end{itemize}
\end{document}